\documentclass[aps,prl,reprint,showpacs]{revtex4-1}

\usepackage{hyperref}
\usepackage{amsmath}
\usepackage{amssymb}
\usepackage{bm}
\usepackage{bbm}
\usepackage{graphicx}
\usepackage{subfigure}
\usepackage{pdfpages}

\begin{document}
\title{Detecting crystal symmetry fractionalization from the ground
  state: Application to $\mathbb Z_2$ spin liquids on the kagome lattice}
\author{Yang Qi}
\affiliation{Institute for Advanced Study, Tsinghua University,
  Beijing 100084, China}
\affiliation{Perimeter Institute for Theoretical Physics, Waterloo, ON
  N2L 2Y5, Canada} 
\author{Liang Fu}
\affiliation{Department of Physics, Massachusetts Institute of
  Technology, Cambridge, MA 02139, USA}

\begin{abstract}
  In quantum spin liquid states, the fractionalized spinon excitations
  can carry fractional crystal symmetry quantum numbers, and this
  symmetry fractionalization distinguishes different symmetry-enriched
  spin liquid states with identical intrinsic topological order. In
  this work we propose a simple way to detect signatures of such
  crystal symmetry fractionalizations from the crystal symmetry
  representations of the ground state wave function. We demonstrate
  our method on projected $\mathbb Z_2$ spin liquid wave functions on the
  kagome lattice, and show that it can be used to classify generic
  wave functions. Particularly our method can be used to distinguish
  several proposed candidates of $\mathbb Z_2$ spin liquid states on the
  kagome lattice.
\end{abstract}
\maketitle

It is well known that anyons in topologically ordered phases can carry
symmetry quantum numbers that are quantized to fractional values.  In
the celebrated example of fractional quantum Hall states, Laughlin
quasi-particles carry fractional charge---the quantum number of the
$U(1)$ symmetry~\cite{Laughlin1983}. In recent years, great progress
has been made in understanding the interplay between symmetry and
fractionalization in other topologically ordered states.  In
particular, topological spin liquids exhibit a more subtle kind of
symmetry fractionalization, associated with the crystal symmetry of
the underlying lattice instead of internal
symmetries~\cite{wenpsg,Essin2013,LuBFU}.  While some aspects of it
have been studied for quite a while, crystal symmetry
fractionalization has now received renewed attention, due to an
increased interest in the role of crystal symmetry in topological
phases of matter. This topic is also becoming timely in view of strong
numerical evidence for spin liquids on kagome lattice found in the
last few years~\cite{Yan2011, Tay2011c, Jiang2012a, Depenbrock2012,
  Rousochatzakis2014}. In order to fully pin down the topological
nature of the numerically found spin liquid liquid, the complete
pattern of crystal symmetry fractionalization needs to be determined.

In this work, we offer a new perspective on crystal symmetry
fractionalization in $\mathbb Z_2$ spin liquids. We find that the nontrivial
way that crystal symmetry acts on an individual anyon is directly
related to the symmetry representation of the topologically ordered
ground states, as labeled by the crystal momentum and parity of
many-body wave functions.  Given that states with different symmetry
labels cannot be adiabatically connected, our finding immediately
makes it clear that the classification of spin liquids is refined and
enriched by taking into account crystal symmetries~\cite{Chen2013}.
Our theoretical result also provides a straightforward method to
classify and detect different spin liquids in numerical studies.  As a
concrete example, we demonstrate that our method can be used to easily
distinguish various $\mathbb Z_2$ spin liquids on the kagome lattice~\cite{sstri,
  wang2007, Lu2011a}.

We begin by briefly reviewing what is known about crystal symmetry
fractionalization in $\mathbb Z_2$ spin liquids, and setting up the
terminology for our work. A $\mathbb Z_2$ spin liquid~\cite{Wen1991,
  Wen1991a} supports three types of anyon excitations: bosonic
spinons, fermionic spinons, and visons. As a defining property of
topological excitations, anyons of each type can only be created in
pairs.  This property makes symmetry fractionalization possible. This
can be understood by considering a many-body excited state containing
two identical anyons that are spatially separated~\cite{Essin2013,
  LuPSGDetect}.  Intuitively speaking, the action of symmetry on this
excited state can then be factorized into a product of two independent
symmetry actions on the anyons.  While the action on a physical state
is necessarily described by a {\it linear} representation of the
symmetry group denoted by $G$, the action on a single anyon is now
allowed to form a {\it projective} representation $\tilde{G}$, such
that the tensor product $\tilde{G} \otimes \tilde{G}$ is a linear
representation of $G$~\cite{Yao2010,Mesaros2013,LuSET,BarkeshliX}.
The projective representation $\tilde{G}$, which has a different group
algebra than $G$, can be regarded as the ``square root'' of $G$. In
this sense, symmetry action on anyons can be called
``fractionalized.'' Throughout this Rapid Communication, a tilde
symbol (\~{}) placed over a symmetry operation means that it acts on
an anyon; otherwise it acts on a physical wave function.

To sharpen the intuitive argument stated above and give a precise
definition of symmetry fractionalization is a nontrivial task that
requires great care.  The main difficulty is that symmetry operations
should in principle be performed on a single anyon, and yet any
physical wave function necessarily contains an even number of
them. (We note that crystal symmetry fractionalization may have
implications for excitation spectra that can be
detected~\cite{wenpsg,Essin2014,WangSET}.) To overcome this
difficulty, we take a different approach and give a precise and {\it
  operational} definition of crystal symmetry fractionalization by
relating it to the \emph{linear} symmetry representations of many-body
wave functions.
  
Crystal symmetries of a given lattice form a space group $G$ generated
by translation, rotation and reflection. Any allowed projective
representation of $G$, denoted by $\tilde{G}$, can be specified by its
modified group algebra as compared to $G$\cite{wenpsg, LuBFU}.  First,
two commuting operations $X$ and $Y$ in $G$, $XY=YX$, can become
anti-commuting in $\tilde{G}$, $\tilde X\tilde Y=-\tilde Y\tilde X$.
This will be referred to as commutation relation
fractionalization. Second, an identity of $X^n=1$ in $G$ can become
$\tilde X^n=-1$ in $\tilde{G}$. This will be referred to as quantum
number fractionalization.

As the main result of this work, we find that the symmetry
representation of ground states is a diagnosis of crystal symmetry
fractionalization in $\mathbb Z_2$ spin liquids. First, the
commutation relation between a translation operation $T_1$ and another
symmetry operation $X$, $\tilde T_1\tilde X=\pm \tilde X\tilde T_1$,
can be determined from the difference between eigenvalues of $X$ for
ground states in \emph{different} topological sectors on a torus
geometry with an \emph{odd} number of unit cells in the direction of
$T_1$.  Second, for an order-two symmetry operation $X^2=1$,
$\tilde X^2=\pm 1$ can be determined from the parity eigenvalue of $X$
acting on ground states on a torus with $4n+2$ sites.  We find it
remarkable that the fractionalized symmetry property of anyons is
simply and directly encoded in the symmetry of ground states.  Some
technical details of our derivation are available in the Supplemental
Material~\footnote{See Supplemental Material for more technical details
  of the derivation discussed in the main text.}.

\paragraph{Fractionalized commutation relation.}
In the presence of a fractionalization in the commutation relation
between $T_1$ and another symmetry operation $X$, ground states in
different topological sectors have different eigenvalues of $X$ on a
torus with odd number of unit cells in the $T_1$
direction~\cite{Essin2013}. To extract the fractionalization of
different anyons, we find it crucial to choose a basis according to
the anyon flux along the direction of $T_1$.

On a torus, a gapped $\mathbb Z_2$ spin liquid has a fourfold ground
state degeneracy, which is protected by its intrinsic topological
order. A basis of these four ground states can be chosen such that
each state carries a different anyon flux going through the torus in
the direction of $T_1$, which can be diagnosed by a Wilson loop
operator in the direction of $T_2$~\cite{Wen1991a}. There are four
types of such flux; each corresponds to one type of anyon excitations
in the toric code topological order. Therefore the four ground states
can be labeled as $|G_a\rangle$, where $a$ denotes the type of anyon.
In this paper we denote the four types of anyon excitations in the
$\mathbb Z_2$ spin liquid state as $a=1$, $b$, $f$, and $v$, standing
for the trivial particle, the bosonic spinon, the fermionic spinon,
and the vison, respectively.

Using this basis, the ratio between parity eigenvalues of two ground
states can be calculated by considering the Berry phase picked up
through the following operations acting on $|G_1\rangle$,
\begin{equation}
  \label{eq:DP}
  X^{-1}(f^a)^{-1}Xf^a|G_1\rangle = e^{i\Delta\Phi}|G_1\rangle,
\end{equation}
where $f^a$ denotes the operation of moving one $a$ anyon across the
torus in the direction of $T_1$ and it maps $|G_1\rangle$ to
$f^a|G_1\rangle=|G_a\rangle$~\cite{kitaev}. This Berry phase
$\Delta\Phi$ can be obtained from the ground state $X$-symmetry
representations as the following,
\begin{equation}
  \label{eq:path}
  \begin{split}
    |G_1&\rangle\xrightarrow{f^a}|G_a\rangle
    \xrightarrow{X}\lambda^a_X|G_a\rangle\\
    &\xrightarrow{(f^a)^{-1}}\lambda^a_X|G_1\rangle
    \xrightarrow{X^{-1}}\lambda^a_X(\lambda^1_X)^{-1}|G_1\rangle,
  \end{split}
\end{equation}
where $\lambda_X^a$ denotes the parity eigenvalue of $X$ acting on
$|G_a\rangle$: $X|G_a\rangle = \lambda_X^a|G_a\rangle$.

On the other hand, the same Berry phase can be obtained using the
projective crystal symmetry representation of the $a$ anyon. Starting
from the ground state $|G_1\rangle$, the operation $f^a$ creates two
anyons locally and moves one across the torus along $T_1$, which is
equivalent to acting $T_1$ on one of the two anyons $n_1$ times, so
the end state can be expressed as $a\otimes\tilde T_1^{n_1}a$, where
$\otimes$ denotes anyon fusion and $n_1$ is the number of unit cells
in the direction of $T_1$, which is an odd number according to our
setup. Then $X$ acts on both anyons and maps the state to
$\tilde Xa\otimes\tilde X\tilde T_1^{n_1}a$. The rest of the actions
can be calculated similarly,
\begin{equation*}
  \label{eq:path2}
  \begin{split}
    &1=a\otimes a\xrightarrow{f^a}a\otimes \tilde T_1^{n_1}a
    \xrightarrow{X}\tilde Xa\otimes \tilde X\tilde T_1^{n_1}a\\
    &\xrightarrow{(f^a)^{-1}}\tilde Xa\otimes \tilde T_1^{-n_1}\tilde
    X\tilde T_1^{n_1}a
    \xrightarrow{X^{-1}}a\otimes \tilde X^{-1}\tilde T_1^{-n_1}\tilde
    X\tilde T_1^{n_1}a.
  \end{split}
\end{equation*}
Hence after the series of operations one anyon is changed into
$\tilde X^{-1}\tilde T_1^{-n_1}\tilde X\tilde
T_1^{n_1}a=(\tau_X^a)^{n_1}a$,
where $\tau_X^a=\pm1$ denotes the fractionalization of the commutation
relation, $\tilde T_1\tilde Xa=\tau_X^a\tilde X\tilde T_1a$.  This can
be simplified as $\tau_X^a$ because $n_1$ is odd and
$(\tau_X^a)^2=1$. Comparing this result with Eq.~\eqref{eq:path} we
obtain the following relation between commutation relation
fractionalization and ground state parity eigenvalues,
\begin{equation}
  \label{eq:etal}
  \tau_X^a=\lambda_X^a / \lambda_X^1.
\end{equation}

\paragraph{Fractionalized quantum number.}
The action of an order-two crystal symmetry $X$ on an anyon is
fractionalized if acting $\tilde{X}$ {\it twice} on a {\it single}
anyon yields $\tilde{X}^2=-1$.  To detect such fractionalized quantum
number, we act $X$ {\it once} on an excited state containing {\it two}
anyons, whose positions are {\it swapped} by $X$ \footnote{Here our
  operational definition of fractionalized quantum number is different
  from Ref.~\onlinecite{LuBFU} but our results are consistent. This is
  detailed in Sec.~I of the Supplemental Material.}. Specifically,
$\tilde X$ maps an anyon at a site $i$ to another anyon at the image
site $X(i)$ and vice versa.

The symmetry action on an anyon is accompanied by additional
gauge transformations~\cite{wenpsg},
\begin{equation}
  \label{eq:Xa}
  \tilde X a_i= U_i a_{X(i)},\quad
  \tilde X a_{X(i)} = U_{X(i)} a_i.
\end{equation}
Therefore, acting $\tilde X$ twice on the anyon $a_i$ leaves anyon at
its original position but yields a factor
$\tilde{X}^2 a = U_{X(i)}U_i a$. Alternatively, if one perform the
operation $X$ once on a physical wave function $|\Psi \rangle$ that
contains a pair of anyons at $i$ and $X(i)$, the same phase factors
$U_i$ and $U_{X(i)}$ are collected from each anyon, so that the pair
of spinons acquires the same total phase of
$U_{X(i)}U_i$. Importantly, if the anyon under consideration is a
fermion, there is an additional statistical sign due to the exchange
of two fermions under $X$. To summarize, when a bosonic anyon carries
a fractionalized quantum number $\tilde X^2=-1$, the parity eigenvalue
of an excited state $|\Psi\rangle$ containing a pair of such anyons is
opposite to that of the ground state, from which $|\Psi\rangle$ is
created.  For fermionic anyons, $\tilde X^2=-1$ implies that the
parity eigenvalues of $|\Psi\rangle$ and $|G\rangle$ are identical.

We now show that detecting $\tilde X^2 = \pm 1$ for spinons can be
further simplified when the spin liquid is constructed from parton
methods (using either the Schwinger-boson or Abrikosov-fermion
approach), which put exactly one spinon on every lattice site.
Specifically, we choose a lattice geometry with $4n+2$ sites, so that
the ground state contains an {\it odd} number of pairs of spinons. The
parity eigenvalue of such a ground state is then equal to the parity
of a single pair of spinons, which in turn detects $\tilde X^2=-1$ as
described above. Importantly, the contribution to the parity
eigenvalue from the fractional quantum number is independent of the
topological sector of the ground state, in contract to the previous
case involving the commutation relation between $\tilde X$ and
$\tilde T_1$. This will be demonstrated with concrete examples below.

\paragraph{$\mathbb Z_2$ spin liquids on kagome lattices.}

We now apply our method of detecting crystal symmetry
fractionalization to $\mathbb Z_2$ spin liquid states on the kagome
lattice.  Recent numerical studies of the spin-$\frac{1}{2}$
Heisenberg model on the kagome lattice ~\cite{Yan2011, Tay2011c,
  Jiang2012a, Depenbrock2012, Rousochatzakis2014} have found strong
evidence for a gapped spin liquid state, likely with a $\mathbb Z_2$
topological order (see however, Ref.~\onlinecite{Iqbal2011}). On the
other hand, various types of $\mathbb Z_2$ spin liquids on the kagome
lattice that differ in symmetry properties have been theoretically
constructed using parton methods in early studies~\cite{sstri,
  wang2007, Lu2011a}.  In what follows, we will connect the numerical
findings to theoretical constructions and show how to determine which
of the $\mathbb Z_2$ spin liquid states theoretically proposed so far
is consistent with the ground state of the Heisenberg model on the
kagome lattice.

To start, we quickly describe the parton construction of various $\mathbb Z_2$
spin liquid states, paying particular attention to the role of crystal
symmetry. Parton constructions postulate that the low-energy dynamics
of the spin liquid phase is described by gapped spinons (which carry
$\mathbb Z_2$ gauge charge) interacting with $\mathbb Z_2$ gauge fields.  Depending on
the $\mathbb Z_2$ background flux patterns, the spinon exhibits different
crystal symmetry fractionalizations, thus leading to distinct spin
liquid states.  This is because in the presence of gauge flux, crystal
symmetries acting on a spinon involve additional $\mathbb Z_2$ gauge
transformations.  For example, when there is a $\pi$ flux within a
unit cell, translations of spinons correspond to the magnetic
translation group with the property
$\tilde{T}_1 \tilde{T}_2 = - \tilde{T}_2 \tilde{T}_1$, resulting in a
fractionalized commutation relation.

In the parton construction, spin liquids with different $\mathbb Z_2$
background flux patterns are classified using the projective symmetry
group (PSG) analysis invented by \citet{wenpsg}, from which we can
derive crystal symmetry fractionalization of bosonic and fermionic
spinons.  Below we derive the crystal symmetry fractionalization for
$\mathbb Z_2$ spin liquids on the kagome lattice that were constructed
using the Schwinger-boson approach in previous works~\cite{sstri,
  wang2007}.  The PSG analysis by \citet{wang2007} has found that
there are four spin liquid states with different $\mathbb Z_2$ flux
patterns, which are adiabatically connected to nearest-neighbor
resonating-valence-bond states and therefore have better variational
energy than other states. Hence we use them as examples to demonstrate
our method of detecting crystal symmetry fractionalization. These four
states are labeled by three $\mathbb Z_2$ variables $(p_1,p_2,p_3)$;
for the sake of completeness, this terminology from
Ref.~\onlinecite{wang2007} is reviewed in Sec.~I of the Supplemental
Material.
  
\begin{figure}[htbp]
  \centering
  \includegraphics{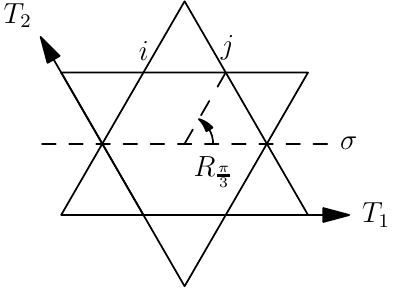}
  \caption{The definition of crystal symmetry operations. $T_{1,2}$,
    $\sigma$, and $R_{\frac\pi3}$ label the symmetry operations of
    translation, mirror reflection, and sixfold rotation,
    respectively.}
  \label{fig:parton}
\end{figure}

The kagome lattice has three independent symmetry operations:
$T_{1,2}$, $\sigma$, and $R_{\pi/3}$, which denote translation, mirror
reflection and rotation, respectively, and their definition is shown
in Fig.~\ref{fig:parton}.  A straightforward translation of
terminology shows that $(p_1, p_2, p_3)$ in the PSG analysis directly
yields the crystal symmetry fractionalizations of the bosonic spinon
excitations (denoted by $b$), listed in the first row of
Table~\ref{tab:csf}.  Here, $\tau_{T_2}= \pm 1$ labels the
fractionalization of the commutation relation between $T_1$ and $T_2$,
defined by $\tilde T_1\tilde T_2= \tau_{T_1} \tilde T_2\tilde T_1$.
Likewise, $\tau_{\sigma}$ is defined by
$\tilde T_1 \tilde \sigma= \tau_{\sigma} \tilde T_1\tilde \sigma$, and
the fractionalization of commutation relation between $T_1$ and the
twofold rotation $R_\pi \equiv R_{\pi/3}^3$ takes the form of
$\tilde T_1\tilde R_\pi=\tau_{R_\pi} \tilde R_\pi\tilde T_1^{-1}$.
All three $\tau$'s are equal for the four spin liquid states we
consider. In addition, quantum number fractionalizations
$\tilde\sigma^2=\pm1$ and $\tilde R_{\pi/3}^6=\pm1$ are listed in the
last two columns of Table II.

\begin{table}[htbp]
  \centering
  \caption{Crystal symmetry fractionalizations of different anyon
    excitations. $b$, $f$, and $v$ denotes the bosonic spinon, the
    fermionic spinon and the vison, respectively.}
  \label{tab:csf}
  \begin{tabular*}{\columnwidth}{@{\extracolsep{\fill}}cccc}
    \hline\hline
    Anyon & $\tau_{T_1}^a=\tau_\sigma^a=\tau_{R_\pi}^a$&
    $\tilde\sigma^2$ & $\tilde R_{\frac\pi3}^6=\tilde R_\pi^2$ \\
    \hline
    $b$ & $(-1)^{p_1}$ & $(-1)^{p_2}$ & $(-1)^{p_1+p_3}$ \\
    $f$ & $(-1)^{p_1+1}$ & $(-1)^{p_2+1}$ & $(-1)^{p_1+p_3+1}$ \\
    $v$ & $-1$ & $+1$ & $+1$ \\
    \hline\hline
  \end{tabular*}
\end{table}

A limitation of the previous PSG analysis is that it is tied to the
Schwinger-boson formalism and hence only gives the crystal symmetry
fractionalization of the bosonic spinons.  The vison excitation in all
four states has the same crystal symmetry
fractionalization~\cite{Huh2011, LuBFU}:
$\tau_{T_2}=\tau_\sigma=\tau_{R_\pi}=-1$ and
$\tilde{\sigma}^2=\tilde R_\pi ^2=1$. This result can be simply
obtained from the charge-flux duality: in a spin liquid state with odd
number of spinons per unit cell as is the case for the kagome lattice,
the vison always sees a $\pi$ flux per unit cell because a spinon is a
$\pi$ flux to a vison. As a result, the vison always has the property
$\tilde{T}_1 \tilde{T}_2 = - \tilde{T}_2 \tilde{T}_1$.

We now use the method of flux attachment to derive the crystal
symmetry fractionalization for the fermonic spinon, which is
equivalent to a composite of a bosonic spinon and a vision---the
latter is a $\pi$ flux to the former. Compared to a bosonic spinon,
the fermionic spinon always sees an extra $\pi$ flux per unit cell due
to the vison attached to it; hence $\tau_{T_2}^b= -\tau_{T_2}^f$.
Similarly, the difference in $\tilde R_\pi^2$ between bosonic and
fermionic spinons follows from the fact that the attachment of a
$\pi$ flux changes the angular momentum between integer and
half-integer values~\cite{Wilczek1982}.  These results can also be
derived using general methods described in Sec~I of the Supplemental
Material, and are summarized in the second row of
Table~\ref{tab:csf}~\footnote{After completing this work we learnt
  that identical results have been obtained in the updated version of
  Ref.~\onlinecite{LuBFU}.}.

\begin{table}[htbp]
  \centering
  \caption{Correspondence between Schwinger-boson and Abrikosov-fermion constructions. $p_i$
    labels the PSG solutions of Schwinger-boson construction~\cite{wang2007}. The $Q_1=\pm
    Q_2$ labels are used by \citet{sstri}, and the labels in the last
    column are used by \citet{Lu2011a}.}
  \label{tab:sbmft}
  \begin{tabular*}{\columnwidth}{@{\extracolsep{\fill}}ccc}
    \hline\hline
    $(p_1,p_2,p_3)$ & Label in Ref.~\onlinecite{sstri}
    & Label in Ref.~\onlinecite{Lu2011a}\\
    \hline
    $(0,0,1)$ & $Q_1=-Q_2$  & Z$_2[0,\pi]\alpha$\\
    $(0,1,0)$ &  $Q_1=Q_2$ & Z$_2[0,\pi]\beta$\\
    $(1,0,1)$ & & Z$_2[0,0]$B\\
    $(1,1,0)$ & & Z$_2[0,0]$A\\
    \hline\hline
  \end{tabular*}
\end{table}

As an independent check of the above results, we find that the above
four spin liquid states constructed from the Schwinger-boson approach
can be equivalently described using the Abrikosov-fermion approach. To
establish the mapping between the two parton constructions, we match
the ground states of spin liquids in the nearest-neighbor
resonating-valence-bond limit, given by the Gutzwiller projection on
the corresponding parton wavefunctions in the two constructions:
 \begin{align}
  \label{eq:swbwf}
  |\psi\rangle&=P_G
  \exp\left[\sum_{\langle ij\rangle}\xi_{ij}
    \epsilon_{\alpha\beta}b_{i\alpha}^\dagger b_{j\beta}^\dagger\right]
  |0\rangle, \\
   \label{eq:swfwf}
  |\psi\rangle&=P_G
  \exp\left[\sum_{\langle ij\rangle}\zeta_{ij}
    \epsilon_{\alpha\beta}f_{i\alpha}^\dagger f_{j\beta}^\dagger\right]
  |0\rangle. 
\end{align}
Here $\xi_{ij}$ and $\zeta_{ij}$ are antisymmetric and symmetric
scalars on the nearest-neighbor bonds $\langle ij\rangle$ used in
Schwinger-boson and Abrikosov-fermion constructions, respectively. The
values of $\xi_{ij}$ or $\zeta_{ij}$ on bonds of the kagome lattice
are given by the corresponding PSG analysis.  By equating
Eq.~\eqref{eq:swbwf} and \eqref{eq:swfwf}, we find explicitly the
mapping between $\xi_{ij}$ and $\zeta_{ij}$ for each of the four spin
liquid states~\cite{Yunoki2006, Yang2012}.  As a byproduct, this
mapping establishes the correspondence between the notation of the
Schwinger-boson~\cite{sstri, wang2007} formalism $(p_1, p_2, p_3)$
with that of the Abrikosov-fermion~\cite{Lu2011a} formalism, as shown
in Table~\ref{tab:sbmft}.  From this mapping, we have confirmed the
results in Table~\ref{tab:csf} (see Sec~II of the Supplementary
Material for details).

\begin{table}[htbp]
  \centering
  \caption{Ratios between $X$-symmetry parity eigenvalues
    of $|G_a\rangle$ and $|G_1\rangle$. The results depend on the
    commutation relation fractionalization $\tau_{T_2}^b=-\tau_{T_2}^f=(-1)^{p_1}$,
    and the ratios are the same for $X=T_2$, $\tau$ and $R_\pi$.}
  \begin{tabular*}{\columnwidth}{@{\extracolsep{\fill}}cccccc}
    \hline\hline
    $p_1$ & $\lambda_X^b/\lambda_X^1$ 
    & $\lambda_X^v/\lambda_X^1$ & $\lambda_X^f/\lambda_X^1$\\
    \hline
    $0$ & $+1$ & $-1$ & $-1$ \\
    $1$ & $-1$ & $-1$ & $+1$ \\
    \hline\hline
  \end{tabular*}
  \label{tab:tras}
\end{table}

Having derived the crystal symmetry fractionalization of anyons for
all four spin liquids on the kagome lattice, we can now determine the
symmetry representations of the ground states for each spin liquid, by
using the general relation between the two as described in the first
part of this work.  First, based on Eq.~\eqref{eq:etal} and the
results of commutation relation fractionalizations summarized in
Table~\ref{tab:csf}, the ratios between symmetry eigenvalues of
different ground states are determined and the results are summarized
in Table~\ref{tab:tras}.

Second, if all other aspects of the symmetry fractionalization are the
same, a difference in the quantum number fractionalization results in
a uniform parity change in the symmetry representation of ground
states in all topological sectors. This relation can be obtained by
explicitly calculating the parity eigenvalues of the model wave
functions in Eq.~\eqref{eq:swbwf} and Eq.~\eqref{eq:swfwf}. The
results are summarized in Table~\ref{tab:tras} and the details of the
derivation is given in Sec.~III of the Supplemental Material. These
results are determined from projected mean field wave functions but
they also apply to general wave functions in the same
topologically ordered phase, because the crystal symmetry
representations are invariant when the state is smoothly deformed
without breaking crystal symmetries.

\begin{table}[htbp]
  \centering
  \caption{Crystal symmetry representations of ground states in
    different topological sectors on a torus with odd-by-$(4n+2)$ unit
    cells. $|G_a\rangle$ denotes the ground state with an anyon flux
    $a$ in the direction of $T_1$, where $a=1$, $b$, $v$, and $f$ denotes the
    trivial anyon, bosonic spinon, the vison and
    the fermionic spinon, respectively.}
  \begin{tabular*}{\columnwidth}{@{\extracolsep{\fill}}cccccc}
    \hline\hline
    $X$ & $|G_1\rangle$ & $|G_b\rangle$ & $|G_v\rangle$ & $|G_f\rangle$\\
    \hline
    $T_2$ & 1 & $(-1)^{p_1}$ & $-1$ & $(-1)^{p_1+1}$\\
    $\sigma$ & $(-1)^{p_2}$ & $(-1)^{p_1+p_2}$
                                        & $(-1)^{p_2+1}$ & $(-1)^{p_1+p_2+1}$\\
    $R_\pi$ & $(-1)^{p_3}$ & $(-1)^{p_1+p_3}$
                                        & $(-1)^{p_3+1}$ & $(-1)^{p_1+p_3+1}$\\
    \hline\hline
  \end{tabular*}
  \label{tab:frac}
\end{table}

Summarizing the above results, we see that the PSG parameters $p_2$
and $p_3$ determines the parity eigenvalues of the ground state
$|G_1\rangle$, and then using $p_1$ the parity eigenvalues of the
other sectors are also determined. Hence we can obtain from
$(p_1,p_2,p_3)$ the crystal symmetry representations of all
topological sectors on a torus with odd-by-$(4n+2)$ unit cells, as
summarized in Table~\ref{tab:frac}.

\paragraph{Conclusions.}
In this work we propose a method to detect crystal symmetry
fractionalizations from the crystal symmetry representations of the
ground states. On a torus with $4n+2$ sites, the ratio between
symmetry parity eigenvalues of ground states in different topological
sectors detects the fractionalization of commutation relation with a
translation symmetry, and a uniform sign change in all sectors detects
the quantum number fractionalization.

Our method can be applied to study the nature of the topological order
of the $\mathbb Z_2$ spin liquid states obtained in numerical
studies. Particularly using the infinite-size density matrix
renormalization group (DMRG) method~\cite{Bauer2014} or an infinite
projected entangled pair states (PEPS) ansatz~\cite{Jordan2008} the
ground states in different topological orders can be obtained on an
infinite cylinder and labeled with the anyon type by calculating the
modular matrices~\cite{Zhang2012, Cincio2013, Bauer2014}. To study
both commutation relation fractionalization and quantum number
fractionalization, we suggest using a cylinder $(4n+2)$ unit cells
wide. Then the crystal symmetry fractionalization studied in this work
can be determined by putting the system on a torus with a length of an
odd number of unit cells and examining the crystal symmetry
representation of topologically degenerate ground states. For the
$\mathbb Z_2$ spin liquid states discussed above, the results are
shown in Table~\ref{tab:frac}.

In this work we explicitly derived the crystal symmetry quantum number
of ground states and the fractional symmetry quantum number of anyons
for four Z2 spin liquid states on the kagome
lattice~\cite{wang2007}. Our method can be used straightforwardly to
study additional spin liquid states that have been theoretically
proposed~\cite{wang2007,Lu2011a,LuBFU}.

\begin{acknowledgments}
  We thank Lukasz Cincio, Michael Hermele, Yuan-Ming Lu, Guifre Vidal,
  Yuan Wan, and Qing-Rui Wang for invaluable discussions.  Y.Q. is
  supported by NSFC Grant No. 11104154. L.F. is supported by the DOE
  Office of Basic Energy Sciences, Division of Materials Sciences and
  Engineering, under Award No. DE-SC0010526. This research was
  supported in part by Perimeter Institute for Theoretical
  Physics. Research at Perimeter Institute is supported by the
  Government of Canada through Industry Canada and by the Province of
  Ontario through the Ministry of Research and Innovation.

  \emph{Note added.} After completing our manucript we were informed
  of a related work~\cite{ZLVPSG}.
\end{acknowledgments}

\bibliography{kpsg,unpublished}

\widetext
\clearpage


\section*{Supplementary material}

# Supplemental Material

In this supplemental material we provide some technical details of the derivations outlined in the main text.

## I. CRYSTAL SYMMETRY FRACTIONALIZATIONS OF BOSONIC AND FERMIONIC SPINONS.

In this section we derive the relation between crystal symmetry fractionalizations of bosonic and fermionic spinons. In a $Z_2$ spin liquid states, both of the three types of topological excitations (bosonic spinon, fermionic spinon and vison) carry projective representations of the crystal symmetry group, and different anyon carries a projective representation with different crystal symmetry fractionalizations. However, the crystal symmetry fractionalizations of different anyons are not independent, and particularly the symmetry fractionalization of fermionic spinon can be derived from the ones of the bosonic spinon and vison, as the former is a bound state of the latter [1]. For $Z_2$ spin liquids with odd number of spinons per unit cell, the symmetry fractionalization of the vison is fixed [2, 3]. Therefore the symmetry fractionalization of the fermionic spinon can be uniquely determined from the one of the bosonic spinon. In this section we focus on the following three crystal symmetry fractionalizations on the kagome lattice: the commutation relation fractionalization between $T_1$ and $T_2$, and the quantum number fractionalization of $\sigma^2$ and $R_\pi^2$. These results are used in the main text to obtain Table I.

Firstly, the fractionalization of commutation relation between $T_1$ and $T_2$ of fermion, $\tau_{T_2}^f$, can be derived from the corresponding fractionalizations of the bosonic spinon and of the vison [1],

$$\tau_{T_2}^f = \tau_{T_2}^b \tau_{T_2}^v. \tag{1}$$

For $Z_2$ spin liquids with odd number of spinons per unit cell, $\tau_{T_2}^v = -1$ [2]. Hence we always have $\tau_{T_2}^f = -\tau_{T_2}^b$ [3]. As explained in the main text this can be understood intuitively using flux attachment.

Next, we consider the reflection symmetry $\sigma$. We find as a bound state of a bosonic spinon and a vison, the fractionalized quantum number of a fermion is the product of the

quantum numbers of a bosonic spinon and that of a vison, times an additional twist factor $t = -1$

$$\tilde{\sigma}_f^2 = -\tilde{\sigma}_b^2 \tilde{\sigma}_v^2, \tag{2}$$

where $\tilde{\sigma}_a$ denotes the projective representation of $\sigma$ acting on anyon type $a$. The twist factor $t = -1$ can be calculated by examining the reflection parity eigenvalue of a wave function $|\Psi\rangle$ containing two fermionic spinon excitations located at two sites related by $\sigma$. As discussed in the main text, if $\tilde{\sigma}_f^2 = +1$, the reflection parity eigenvalues of $|\Psi\rangle$ and of the ground state $|G\rangle$ are opposite because of the Fermi statistical sign of exchanging two fermions, and if $\tilde{\sigma}_f^2 = -1$ the two eigenvalues are the same. On the other hand, $|\Psi\rangle$ can also be viewed as a wave function containing two bosonic spinons and two visons. Therefore the ratio between reflection parity eigenvalues of $|\Psi\rangle$ and of $|G\rangle$ is also given by $\tilde{\sigma}_b^2 \tilde{\sigma}_v^2$. Comparing these two results we conclude that the twist factor $t = -1$ in Eq. (2). For the $Z_2$ spin liquid states we consider, the vison always has $\tilde{\sigma}_b^2 = +1$ [2]. Hence the fermionic spinon and bosonic spinon always have opposite reflection quantum number fractionalization $\tilde{\sigma}_f^2 = -\tilde{\sigma}_b^2$.

Our conclusion of $t = -1$ should be contrasted with the incorrect result of the twist factor obtained in Ref. 1. We note that the phase of the two-fermion wave function depends on the ordering of the two fermions, and after mirror reflection the exchange of the two fermions gives the Fermi statistical sign of $-1$. This minus sign is the key in the above derivation of $t = -1$. As a independent check, in Sec. II we explicitly derive the mapping between Schwinger-boson and Abrikosov-fermion constructions, which gives the same mapping between the crystal symmetry fractionalizations of the bosonic and fermionic spinons.

We also note that a different operational definition was used by Lu *et al.* [3], which lead to the identical result. In their definition they considered two anyons, whose positions are *unchanged* under mirror reflection, in contrast to our definition where the position of the anyons are *exchanged* by mirror reflection. Their definition is more involved to study when the translation operator (which connects the two positions of the anyons) does not commute with the reflection.

Similarly for the inversion symmetry the twist factor $t$ is also $-1$ because of the Fermi statistical sign, so $\tilde{R}_{\pi,f}^2 = -\tilde{R}_{\pi,b}^2 \tilde{R}_{\pi,v}^2$. Again for the $Z_2$ spin liquids on the kagome lattice we always have $\tilde{R}_{\pi,v}^2 = +1$. This twist factor $t = -1$ was also obtained using different methods by Essin and Hermele [1]. Therefore the fermionic spinon and bosonic spinon have opposite inversion quantum number fractionalization $\tilde{R}_{\pi,f}^2 = -\tilde{R}_{\pi,b}^2$.

## II. ABRIKOSOV-FERMION REPRESENTATION OF SCHWINGER-BOSON SPIN LIQUID STATES

In this section we discuss the Abrikosov-fermion representation of the nearest-neighbor Schwinger-boson spin liquid states. In general these two ways of constructing spin liquid states can describe several common $Z_2$ spin liquid states [3], and particularly spin liquid states with pairings on nearest-neighbor bonds only can be described by both of the two constructions [4, 5].

We begin with a brief review of the four bosonic PSG solutions discussed in the main text, which were studied in details by Wang and Vishwanath [6]. These four PSG solutions are labeled by three $Z_2$ variables $(p_1, p_2, p_3)$, which parameterizes the projective crystal symmetry representation carried by the bosonic spinon operators,

$$\phi_{T_1}(\boldsymbol{r}) = 0, \tag{3a}$$

$$\phi_{T_2}(\boldsymbol{r}) = p_1 \pi r_1, \tag{3b}$$

$$\phi_\sigma(\boldsymbol{r}) = p_2 \frac{\pi}{2} + p_1 \pi r_1 r_2, \tag{3c}$$

$$\phi_{\frac{\pi}{3}}(\boldsymbol{r}) = p_3 \frac{\pi}{2} + p_1 \pi \frac{r_2(r_2 - 1 + 2r_1)}{2}. \tag{3d}$$

where $\phi_X(\boldsymbol{r})$ denotes the additional U(1) phase acquired by the bosonic operator after the symmetry operation $X$. In other words, the projective symmetry representation of $X$ carried by $b_{i\alpha}$ is

$$X: b_{i\alpha} \to e^{i\phi_X(\boldsymbol{r}_i)} b_{X(i)\alpha}. \tag{4}$$

The PSG solution in Eq. (3) determines both the crystal symmetry fractionalization of the bosonic spinon and the pattern of $\xi_{ij}$ in the mean field wave function in Eq. (6) of the main text. On one hand, the PSG solution describes the projective representation of the crystal symmetry group carried by the bosonic spinons, and from it their crystal symmetry fractionalization listed in Table I can be derived. On the other hand, it also describes the pattern of $\xi_{ij}$ such that the wave function in Eq. (6) is invariant under symmetry transformations. The patterns of $\xi_{ij}$ for the four different PSG solutions are plotted in Fig. 1, where on each bond the value of $\xi_{ij} = \pm 1$ is represented by an arrow because $\xi_{ij} = -\xi_{ji}$. These patterns are the same as the patterns of the nearest-neighbour pairing terms listed in Ref. 6.

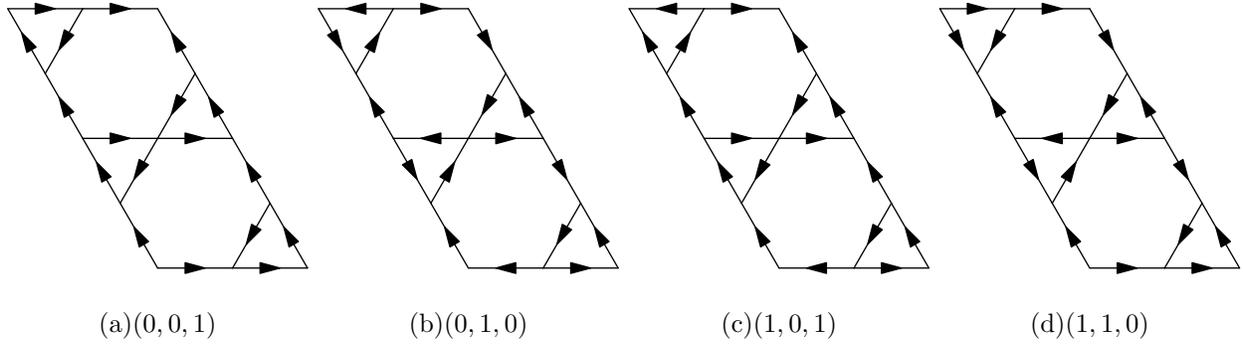

FIG. 1. Pairing amplitude $\xi_{ij}$ of Schwinger-boson mean field constructions with four different PSG solutions. The arrows represent the value of $\xi_{ij} = \pm 1$, where an arrow from $i$ to $j$ indicates $\xi_{ij} = -\xi_{ji} = +1$.

TABLE I. Summary of PSG solutions for the Schwinger-boson and Abrikosov-fermion construction of the four $Z_2$ spin liquid state.

| $(p_1^b, p_2^b, p_3^b)$ | $(p_1^f, p_2^f, p_3^f)$ |
|---|---|
| $(1,1,0)$ | $(0,0,0)$ |
| $(0,1,0)$ | $(1,0,0)$ |
| $(1,0,1)$ | $(0,1,1)$ |
| $(0,0,1)$ | $(1,1,1)$ |

Starting from a specific Schwinger-boson PSG solution, we derive the PSG solution of the corresponding Abrikosov-fermion PSG by first working out the pattern of $\zeta_{ij}$ using the correspondence between $\xi_{ij}$ and $\zeta_{ij}$ given by Ref. 5, and then identify the compatible

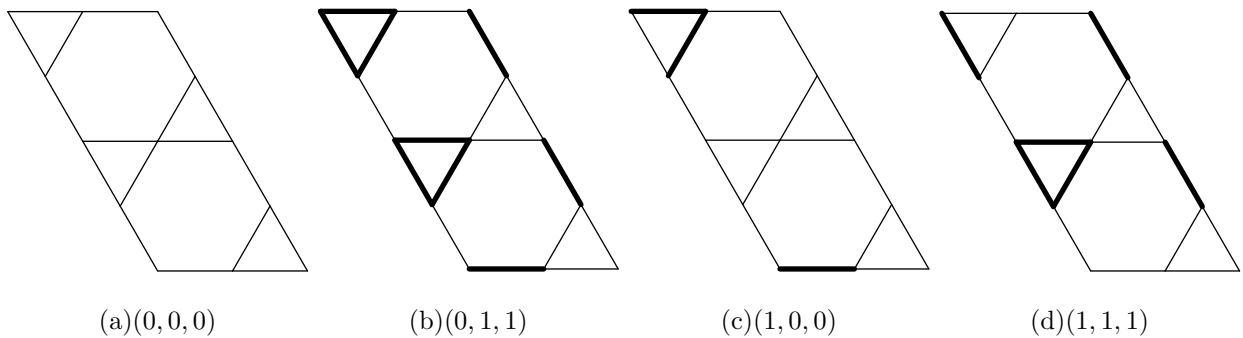

FIG. 2. Pairing amplitude $\zeta_{ij}$ of Abrikosov-fermion mean field constructions with four different PSG solutions. A thin (thick) bond represents $\zeta_{ij} = +1$ (-1), respectively.

PSG solutions, which are described by the other four combinations of $(p_1, p_2, p_3)$ that does not describe a nearest-neighbor Schwinger-boson ansatz. These Abrikosov-fermion PSG solutions are a subset of the ones given by the general classification in Ref. 7. Here we denote the PSG parameters of the Schwinger-boson and Abrikosov-fermion ansatz by $p_i^b$ and $p_i^f$, respectively. For each Schwinger-boson state labeled by $(p_1^b, p_2^b, p_3^b)$, the corresponding Abrikosov-fermion PSG solution $(p_1^f, p_2^f, p_3^f)$ and the label as used in Ref. 7 is listed in Table I.

In these four Abrikosov-fermion PSG solutions, the fermionic spinons transform under crystal symmetry operations with an additional U(1) phase described by Eq. (3), and these solutions are a subset of the possible PSG solutions, with the general SU(2) gauge transformations restricted to the U(1) subgroup. Hence these four states are included in the general classification given by Lu *et al.* [7]. We can label these states with the notation used in Ref. 7 by calculating the topological invariants of symmetry fractionalizations listed in Ref. 3.

TABLE II. Summary of PSG solutions and exactly solvable Hamiltonians of the four different $Z_2$ state realizable in the quantum dimer model.

| Symmetry fractionalizations | $(p_1^f, p_2^f, p_3^f)$ | SU(2) PSG |
|---|---|---|
| $T_2^{-1} T_1^{-1} T_2 T_1$ | $(-1)^{p_1^f}$ | $\eta_{12}$ |
| $\sigma^2$ | $(-1)^{p_2^f}$ | $\eta_\sigma \eta_{\sigma C_6}$ |
| $(R_{\frac{\pi}{3}} \sigma)^2$ | $(-1)^{p_2^f + p_3^f}$ | $\eta_\sigma$ |
| $(R_{\frac{\pi}{3}})^6$ | $(-1)^{p_1^f + p_3^f}$ | $\eta_{C_6}$ |
| $\sigma^{-1} T^{-1} \sigma T$ | $(-1)^{p_2^f}$ | $\eta_{\sigma T} \eta_{C_6 T}$ |
| $R_{\frac{\pi}{3}}^{-1} T^{-1} R_{\frac{\pi}{3}} T$ | $(-1)^{p_3^f}$ | $\eta_{C_6 T}$ |

TABLE III. Summary of PSG solutions and exactly solvable Hamiltonians of the four different $Z_2$ state realizable in the quantum dimer model.

| $(p_1^f, p_2^f, p_3^f)$ | $\eta_{12}$ | $\eta_\sigma$ | $\eta_{\sigma T}$ | $\eta_{\sigma C_6}$ | $\eta_{C_6 T}$ | $\eta_{C_6}$ | SU(2) PSG Label |
|---|---|---|---|---|---|---|---|
| $(0,0,0)$ | $+1$ | $+1$ | $+1$ | $+1$ | $+1$ | $+1$ | $Z_2[0,0]A$ |
| $(1,0,0)$ | $-1$ | $+1$ | $+1$ | $+1$ | $+1$ | $-1$ | $Z_2[0,\pi]\beta$ |
| $(0,1,1)$ | $+1$ | $+1$ | $+1$ | $-1$ | $-1$ | $-1$ | $Z_2[0,0]B$ |
| $(1,1,1)$ | $-1$ | $+1$ | $+1$ | $-1$ | $-1$ | $+1$ | $Z_2[0,\pi]\alpha$ |

Using the PSG solutions in Eq. (3), six symmetry fractionalization invariants are calcu-

lated and listed in Table II. These results are then compared to the same invariants expressed using the $Z_2$ labels $\eta_A$ for generic SU(2) PSG solutions[3, 7]. Using these relations we can cauculate the labels $\eta_A$ for the four PSG solutions labeled by $p_i^f$ and the results are listed in Table III. Finally these states can be labeled by the notations defined by Lu *et al.* [7] by comparing the results of $\eta_A$ to Table III in Ref. 3, and these labels are listed in the last column of Table III.

In summary, starting from a nearest-neighbor pairing Schwinger-boson state with a certain PSG solution, we can first derive the corresponding fermionic PSG solution labeled by $(p_1^f, p_2^f, p_3^f)$ using the mapping in Table I, then label the PSG solution using the notation in Ref 7 using the mapping in Table III. These results are presented in Table II in the main text.

### III. PARITY EIGENVALUES OF THE GROUND STATE WAVE FUNCTIONS.

In this section we calculate the $\sigma$ and $R_\pi$ parity eigenvalues of the ground state wave functions, which is used to detect the fractionalized quantum numbers of $\tilde{\sigma}^2$ and $\tilde{R}_\pi^2$, as explained in the main text. We first obtain the results using the model wave functions in Eq. (6) and Eq. (7) in the main text, then argue that this result also applies to generic wave functions that belong to the same SET phase.

Here we put our system on a torus with $4n + 2$ sites, so there must be an odd number of unit cells along one direction, and as shown in Table III in the main text different ground states have different parity eigenvalues of $\sigma$ and $R_\pi$. Hence we also need to determine the topological sector the mean field wave functions belong to. Using the sign structure in the mean field wave functions one can show that [8] the Schwinger-boson wave function in Eq. (6) in the main text is a superposition of $|G_1\rangle$ and $|G_b\rangle$, while the Abrikosov-fermion wave function in Eq. (7) in the main text is a superposition of $|G_v\rangle$ and $|G_b\rangle$. Using this result it is easy to identify the ground state crystal symmetry representations from the projected mean field wave functions.

For the two states with $p_1^b = 0$, the crystal symmetry representations can be determined using the Schwinger-boson wave function. In this case their PSG solutions in Eq. (3) are simplified to that after the action of $\sigma$ or $R_\pi$ the spinon operator $b_{i\alpha}$ acquires a phase of $p_2^b \frac{\pi}{2}$ and $p_3^b \frac{\pi}{2}$, respectively [6]. Therefore the wave function, containing $4n + 2$ spinons after

the Gutzwiller projection, acquires a phase of $p_2^b\pi$ and $p_3^b\pi$, respectively. Since this wave function is a superposition of $|G_1\rangle$ and $G_b\rangle$, we conclude that their parity eigenvalues under $\sigma$ and $R_\pi$ is $(-1)^{p_2^b}$ and $(-1)^{p_3^b}$, respectively.

For the two states with $p_1^b = 1$, we need to use the Abrikosov-fermion wave function. (In this case the Schwinger-boson wave function is not symmetric because it is a superposition of $|G_1\rangle$ and $|G_b\rangle$, which has different crystal symmetry representations in this certain geometry according to Table III in the main text.) Similarly in these two states the PSG solutions also simplies to that after the action of $\sigma$ or $R_\pi$ the spinon operator $f_{i\alpha}$ acquires a phse of $p_2^f \frac{\pi}{2}$ and $p_3^f \frac{\pi}{2}$, respectively. However in this case in addition to this phase from each spinon, there is an additional sign from the reordering of fermion operators in the Gutzwiller projection operator after the symmetry transformation, which is worked out in Sec. IV. Combing these two phase factors we conclude that the parity eigenvalues of the Abrikosov-fermion wave function, which is a superposition of $|G_v\rangle$ and $|G_b\rangle$, is $(-1)^{p_2^f} = (-1)^{p_2^b+1}$ and $(-1)^{p_3^f+1} = (-1)^{p_3^b+1}$ under $\sigma$ and $R_\pi$, respectively (we used the mapping between $p_i^b$ and $p_i^f$ in Table I). Hence according to the results in Table III in the main text the parity eigenvalues of $|G_1\rangle$ are $(-1)^{p_2^b}$ and $(-1)^{p_3^b}$.

## IV. FERMION REORDERING SIGNS.

In this section we compute the fermion operator reordering signs that appears in the definition of the Gutzwiller projection operator for the Abrikosov-fermion mean field state in Eq. (7) in the main text. The projection operator $P_G$ contain a product of fermion operators on each lattice sites, and after a crystal symmetry transformation the ordering of the sites is changed, so the aforementioned wave function can acquire an additional minus sign due to the reordering of fermion operators. This sign $\phi_X^f$, where $X$ denotes the crystal symmetry operation, is computed here, for a system with $4n + 2$ sites.

For the translation symmetry operation $T_1$, we can order the fermion operators first from left to right in each row in the direction of $T_1$, then from top to bottom for each row in the direction of $T_2$. After translation, in each row we need to move one fermion operator from the beginning to the end. Assuming there are $n_1$ and $n_2$ unit cells in the direction of $T_1$ and $T_2$, respectively, the reordering of the fermion operators gives a sign of $(-1)^{n_1+1}$ for each row, and a total sign of $(-1)^{n_2(n_1+1)} = (-1)^{n_2}$ because the total number of unit cells $n_1 n_2$

is odd. Since we have an odd number of sites per unit cell and each sublattice is giving the same contribution, the total phase is $\varphi^f_{T_1} = n_2\pi$. Similarly we have $\varphi^f_{T_2} = n_1\pi$. Note that since $n_1 n_2 = 4n + 2$, one of these two factors is $\pi$ and the other is zero.

For the inversion symmetry operation $R_\pi$, we can order the fermion operators in a way that the two sites related by $R_\pi$ appear together:

$$|\psi\rangle_p = \left(f^\dagger_{i_1} f^\dagger_{i'_1}\right)\left(f^\dagger_{i_2} f^\dagger_{i'_2}\right) \cdots \left(f^\dagger_{i_{6n+3}} f^\dagger_{i'_{6n+3}}\right)|0\rangle, \tag{5}$$

where $f^\dagger_i = \sum_a f^\dagger_{ia}$, and $i^\dagger_l$ labels the lattice site that is related by $R_\pi$ to site $i$. Since there are in total $4n+2$ unit cells in the system, there are in total $6n+3$ pairs of sites. After the operation of $R_\pi$, we need to exchange the two fermion operators within each pair to restore the fermion order, and each pair contributes a minus sign. Since the total number of pairs is odd, we conclude $\varphi^f_{R_\pi} = \pi$.

For the mirror reflection symmetry $\sigma$, we also order the fermions such that two sites related by $\sigma$ appear together. However, as shown in Fig. 1 in the main text, there are also sites that are cut by the mirror plane and stays invariant under mirror reflection, and we put them in the beginning of the product, as the following,

$$|\psi\rangle_p = f^\dagger_{j_1} \cdots f^\dagger_{j_p} \left(f^\dagger_{i_1} f^\dagger_{i'_1}\right)\left(f^\dagger_{i_2} f^\dagger_{i'_2}\right) \cdots \left(f^\dagger_{i_q} f^\dagger_{i'_q}\right)|0\rangle, \tag{6}$$

where $j_1, \ldots j_p$ denotes the sites that are on the mirror axis $\sigma$. The total number of such sites $p$ equals to the number of unit cells along the direction of $\sigma$. In our setup, the number of sites on the mirror plane $\sigma$ always equal to two modular four. Hence the number of pairs $q$ in Eq. (6) must be an even number. Similar to the previous discussion we conclude $\varphi^f_\sigma = 0$.

---



\end{document}